\documentclass[a4paper]{JHEP3}
\usepackage{epsfig}
\usepackage{graphicx}
\usepackage{amssymb}
\usepackage{latexsym}

\newcommand{\be}{\begin{equation}}
\newcommand{\ee}{\end{equation}}

\title{Optimal weak measurements: Mixed States.}
\author{N.D. Hari Dass\footnote{dass@tifrh.res.in}
\\  TIFR-TCIS, Hyderabad 500046.\\}
\author{Rajath Krishna R\footnote{rajathkrishnar@gmail.com}\\
 Imperial College, London. } 

\abstract{In an earlier publication we had given an exhaustive analysis of the criteria for weak value measurements of pure states to
be optimal in the sense considered by Wootters and Fields. We had proved, for arbitrary spin cases, that the measurements are optimal
when the post-selected state is mutually unbiased wrt the eigenstates of the observable being measured.Here we extend the discussion to
mixed states. For these, weak value measurements have several problems which we illustrate with the protocol proposed by Shengjun Wu. We
discuss tomography of mixed states based on weak measurements and show that while the principal results of Wootters and Fields hold, namely,
the set of observables needed for complete tomography are such that their eigenstates form a mutually unbiased bases, weak tomography removes
a serious lacuna from the Wootters and Fields analysis i.e the need to consider only state averaged error volumes or information.
We also consider another proposal for weak tomography of mixed states by Lundeen and Bamber, and reach similar conclusions about MUB. 
}


\begin{document}

\section{Standard tomography of mixed states}
Consider states(density matrices) operating on a N-dimensional Hilbert space. Being traceless and Hermitian, they require $N^2\,-\,1$
{\em real} parameters for their complete specification. An important subset of them, {\em the pure states} require $2(N\,-\,1)$ real
parameters. The latter always form even-dimensional manifolds and are in fact globally complex manifolds. The real dimensionality of the former
can be both odd and even, and hence these can not always be complex manifolds.

An important issue in quantum theory is that of {\em tomography} i.e determination of the state through suitable measurements. In what one
may call standard tomography, one measures the expectation values of $N^2\,-\,1$ suitable observables(Hermitian) in the state under
consideration, and deduces the density matrix.

To begin with one expands the Hermitian density matrix in a {\em basis} of $N^2\,-\,1$ {\em traceless Hermitian} operators $T_i$:
\be
\label{density}
\rho\,=\,\frac{I}{N}\,+\,\sum_{i=1}^{N^2\,-\,1}\,c_i\,T_i
\ee
Defining
\be
\label{rhobasis}
{\cal M}_{ij}\,=\,tr\,T_i\,T_j
\ee
and the expression for expectation values of operators
\be
\label{expectations}
\langle\,O\,\rangle_\rho\,=\,tr\,\rho\,O
\ee
one gets 
\be
\label{tomography}
x_j\,=\,\sum_i\,{\cal M}_{ji}\,c_i
\ee
where $x_j\,=\,tr\,\rho\,T_i$. This can be solved to determine $c_i(x)$ as functions of $x_i$.

Wootters and Fields \cite{woottersmub} addressed the optimality issue for such tomography by determining the conditions under which the {\em error
volumes} are the least. This is done through the introduction of a metric on the space of $\rho$'s.

\be
\label{metric}
dl^2\,=\,2\,tr\,d\rho\cdot d\rho
\ee
where 
\be
\label{rhodiff}
d\rho\,=\,\sum_i\,dc_i\,T_i
\ee
This way one can obtain the metric on state space in terms of the coordinates $x_i$. It is clear from eqn.(\ref{metric}) that
the space of mixed states is {\em flat}. For pure states, on the other hand, the requirement $\rho^2\,=\,\rho$ defines a
hypersurface for which the metric is nontrivial.

The extents of the {\em error parallelopiped} were taken by Wootters and Fields to be the respective variances. These are
generically state-dependent.Though in their analysis the metric itself was state-independent, the state dependence of the
error volumes necessitated a state-averaging for obtaining optimality criteria.

Their principle result was that the measurements are optimal when the set of observables needed for tomography are such that their 
eigenvectors form {\em mutually unbiased bases(MUB)}, a poerful concept first introduced by Schwinger \cite{schwingermub}.

At this stage what is not specified is how the expectation values $x_i$ are measured. For example, they could be measured
through the so called projective or strong measurements in which case the errors are just the quantum mechanical variances
associated with the state. In such measurements definite values of the observable can be associated with the pointer positions
of the apparatus. The variance in the pointer positions equals the quantum mechanical {\em uncertainties} of the observables in
question and are state dependent. Furthermore, the state after measurements is the eigenstate corresponding to the eigenvalue 
associated with the pointer position, and is hence 'far' from the state before measurement.

\section{Weak tomography of mixed states-I}
In contrast, when the expectation values are obtained by the so called {\em weak measurements} as pioneered by Aharonov and his
collaborators \cite{aharorig} (we shall consider {\em weak value} measurements shortly), the pointer reading of the apparatus can not be 
associated with any value of the observable. Not only is the state of the system not any eigenstate of the measured observable, it
is actually only very slightly different from the pre-measurement state. Nevertheless, the average pointer position even in this
type of measurements is the expectation value of the observable in the state under consideration!

Therefore, if expectation values of the optimal set of observables can be determined this way, state tomography can again be
realised. In as far as the determination of the density matrix in terms of the observed expectation values is concerned, it works
exactly as the standard tomography is concerned. But for optimality criterion, there is an important difference! The variance in
weak measurements, with or without post-selection, are essentially determined by the measurement noise, which is very large. The
state dependence of the variance is extremely weak. This makes the volume of the error parallelopiped to be also state
independent and optimality criterion can now be obtained without {\em state averaging}! This is an important improvement.

Of course, all the generic disadvantages of weak easurements like the necessity of very very large ensemble sizes to
compensate for the large noises etc have to be reckoned with.

So the same conclusions as were reached by Wootters and Fields regarding the MUB criterion will hold now too.

\section{Lundeen - Bamber scheme for weak tomography of mixed states}
It is clear that no matter which strategy one uses, one must get $N^2\,-\,1$ {\em real} data for complete state determination. In
the method discussed so far these came from the expectation values of that many {\em different} observables. In the weak value
tomography of pure states, one makes measurements of a {\em single} observable and use a particular pure state for post-selection. To
be more precise, one measures the $N\,-\,1$ independent {\em Projections} onto the eigenstates of the observable yielding $N\,-\,1$ 
{\em complex} parameters which is precisely the data required to specify pure states.

Lundeen and Bamber \cite{lu2} have prescribed a method for weak tomography of mixed states which superficially resembles weak value 
tomography, but with crucial differences. The common features are that only a single observable is made use of (ostensibly) and
a {\em known} pure state $|b\rangle$ is used in addition. But the main differences are i) there is no post-selection in the sense that the post-measurement
state is not fixed, and, ii) given the observable to be A and its eigenstates $|a_i\rangle$, instead of the projections 
$|\Pi_i=|a_i\rangle\langle a_i|$ used in the weak value measurements of pure states, \cite{lu2} use the $N^2\,-\,N$ 
{\em non-hermitian operators} $O_{ij}, j\,\ne\,i$ given by
\be
\label{lundeenops}
O_{ij}\,=\,\Pi_i\,\Pi_b\,\Pi_j\quad\quad \Pi_i\,=\,|a_i\rangle\langle a_i|\quad\quad \Pi_b\,=\,|b\rangle\langle b|
\quad\quad c_i\,=\,\langle\,b|a_i\rangle\,\ne\,0
\ee
along with N hermitian projectors $O_{ii}$. Because of the relations
\be
\label{lundeenops2}
\sum_i\,\frac{O_{ii}}{|c_i|^2}\,=\,I\quad\quad O_{ij}^\dag\,=\,O_{ji}
\ee
there are exactly $N^2\,-\,1$ hermitian operators in the set and their expectation values in the state $\rho$ to be determined are 
\be
\label{lundeenexp}
\langle\,O_{ij}\,\rangle_\rho\,=\,tr\,\rho\,O_{ij}\,=\,\rho_{ji}\,c_i^*\,c_j\quad\quad \rho_{ji}\,=\,\langle a_j|\rho|a_i\rangle
\ee
When these are determined through weak measurements, we shall denote them by $w_{\rho,ij}$, but these are not weak values. They satisfy
\be
\label{wijreality}
w_{\rho,ij}\,=\,\rho_{ji}\,c_i^*\,c_j\quad\quad \rho_{ji}\,=\,\langle a_j|\rho|a_i\rangle
\ee
As $w_{ij}$ are directly proportional to $\rho_{ji}$ \cite{lu2} call this a {\em direct} determination of the state, but it should be emphasised
that even standard tomography is 'direct' in this mathematical sense! This is the story as far as tomography is concerned.

But as can be understood from \cite{ourpure,woottersmub}, optimality criterion requires a knowledge of the metric and volume element on the
state space. It is clear that the independent components of the density matrix can themselves be chosen as the coordinates for this space.
Then, eqn.(\ref{metric}) will induce a metric on the state space expressed in these coordinates. Expanding
\be
\label{metric2}
dl^2\,=\,\sum_i\,\rho_{ii}^2\,+\,\sum_{i<j}\,|\rho_{ij}|^2
\ee
This is a positive semidefinite quadratic form with constant coefficients, but the $N^2$ $\rho_{ij}$ are not all independent because of the 
$tr \rho\,=\,1$ constraint. This is a linear constraint only involving the diagonal elements of $\rho$. We can eliminate any diagonal element
in terms of the other $N\,-\,1$. This will still leave eqn.(\ref{metric2}) as a positive semi-definite quadratic form albeit non-diagonal. But
the coefficients are still constant, representing a flat metric. The determinant of this metric is just a number, possibly dependent on N, say,
$d(N)^2$.

Now let us turn to using the weak expectation values $w_{\rho,ij}$ as coordinates. They too are not all linearly independent, but the
independent ones can be found by simply scaling the independent components of $\rho$ according to eqn.(\ref{wijreality}). To determine the 
optimality criterion we only need the determinant $g$ of the metric on state space in these weak coordinates; but that is trivial to determine
\be
\label{gdirect}
g\,=\,\frac{d(N)^2}{(\prod_{i=1}^N\,|c_i|^2)^2}
\ee
The corresponding volume element is
\be
\label{volumedirect}
dV\,=\,\frac{d(N)}{(\prod_i\,|c_i|^2)}\,\prod_{j=1}^{(N^2\,-\,1)}\,dx_j
\ee
where we have denoted the coordinates by $x_j$. With weak measurements, the error in every coordinate direction is set by the noise $\Delta_w$
which has a very mild dependence on state that can be safely neglected. Thus the volume of the error parallelopiped is given by
\be
\label{errvol}
\Delta\,V_{err}\,=\,
\frac{d(N)\,\Delta_w^{N^2\,-\,1}}{(\prod_i\,|c_i|^2)}
\ee
The most important features of this error volume are twofold: i) it is {\em state-independent}, unlike the case in \cite{woottersmub},
and, ii) as a consequence it is valid for arbitrary size errors.

The state independence has the important consequence that the error volume can be minimised without the need for any state averaging,
as long as one takes into account the constraint
\be
\label{bnorm}
\sum_{i=1}^N\,|c_i|^2\,=\,1
\ee
yielding the optimality conditions
\be
\label{mixeddirectoptimal}
|c_i|^2\,=\,\frac{1}{N}
\ee
i.e, the state $|b\rangle$ is mutually unbiased wrt the eigenstates $|a_i\rangle$.
\section{Weak value measurements for mixed states}
In contrast to using weak measurements for tomography, the so called {\em weak values} can also be used for tomography. We call this 
{\em weak value tomography} to distinguish the case discussed earlier. For pure states these were pioneered by \cite{lu1,swutomo}. 
Shengjun Wu \cite{swutomo} has also proposed a scheme for weak value measurements for mixed states. He considers the $N\,-\,1$ independent
projections $\Pi_i\,=\,|a_i\rangle\langle a_i|$, but to get sufficient data for mixed state tomography, he employs several post-selection
states.

The weak value of an observable S in the mixed state $\rho$ when post selected by $|b\rangle$ is given by
\be
\label{wvmixed}
w_S\,=\,\frac{tr\,\rho\,\Pi_b\,S}{tr\,\rho\,\Pi_b}
\ee
Applying this to the projectors $\Pi_i\,=\,|a_i\rangle\langle a_i|$ and the post-selected states $|b_j\rangle$, one gets
\be
\label{swuwv}
w^\rho_{ji}\,=\,\frac{\langle a_i|\rho_|b_j\rangle\,\langle b_j|a_i\rangle}{\langle b_j|\rho|b_j\rangle}
\ee
To see implications for tomography, \cite{swutomo} reexpress this in two equivalent forms:
\be
\label{swu1}
\langle a_i|\rho|a_j\rangle\,=\,\sum_k\,P_k\,\frac{\beta_{kj}}{\beta_{ki}}\,w^{\rho}_{ki}\quad\quad P_k\,=\,\langle b_k|\rho|b_k\rangle\quad\quad \beta_{kj}\,=\,\langle\,b_k|a_j\rangle
\ee
and,
\be
\label{swu2}
\langle\,b_i|\rho|b_j\rangle\,=\,P_j\,\sum_k\,\frac{\beta_{ik}}{\beta_{jk}}\,w^{\rho}_{jk}
\ee
There are two {\em apparent} difficulties for tomography this way. In addition to the weak values $w^{\rho}$, one has to measure the probabilities $P_k$ also. This will complicate the error analysis, though this can be circumvented by using the second form and introducing 
\be
\label{swutemp}
x_{ij}\,=\,\frac{\langle\,b_i|\rho|b_j\rangle}{P_j}
\ee
But the most serious problem with this approach is that while $N^2\,-\,1$ {\em real} values are needed for complete tomography, here we have
$(N\,-\,1)\cdot M$ {\em complex} data(M is the number of post-selected states used). A match is possible only if M exactly equals 
$\frac{(N\,+\,1)}{2}$. For even N, there can never be a match. For odd N, while a match is possible, the type of analysis used in 
\cite{ourpure} is not
easy to come up with.

\acknowledgments
RK thanks TCIS-Hyderabad for the hospitality during which this work was initiated. 

\end{document}